\documentclass[11pt,twoside]{article}
\usepackage{asp2010}
\usepackage{epsf}
\usepackage{lscape}
\resetcounters

\aspvoltitle{Progress in Physics of the Sun and Stars: A New Era in Helio- and Asteroseismology}
\aspvolauthor{H. Shibahashi and A.E. Lynas-Gray, eds. }
\aspvolume{479}
\aspcpryear{2013}
\begin{document}
\setcounter{page}{395}

\title{Helioseismic Constraints and Paradigm Shift in Solar Dynamo}
\author{Alexander~G.~Kosovichev$^1$, Valery~V.~Pipin$^{1,2,3}$, and Junwei~Zhao$^1$
 \affil{$^1$Stanford University, Stanford, CA 94305, USA}
 \affil{$^2$UCLA, Los Angeles, CA 90095, USA}
 \affil{$^3$Institute of Solar-Terrestrial Physics, Irkutsk, 664033, Russian Federation}}

\begin{abstract}
Helioseismology provides important constraints for the solar dynamo
problem. However, the basic properties and even the depth of the
dynamo process, which operates also in other stars, are unknown.
Most of the dynamo models suggest that the toroidal magnetic field
that emerges on the surface and forms sunspots is generated near the
bottom of the convection zone, in the tachocline. However, there is
a number of theoretical and observational problems with justifying
the deep-seated dynamo models. This leads to the idea that the
subsurface angular velocity shear may play an important role in the
solar dynamo. Using helioseismology measurements of the internal rotation and meridional circulation, we investigate a mean-field MHD model of dynamo
distributed in the bulk of the convection zone but shaped in a
near-surface layer. We show that if the boundary conditions at the
top of the dynamo region allow the large-scale toroidal magnetic
fields to penetrate into the surface, then the dynamo wave
propagates along the isosurface of angular velocity in the
subsurface shear layer, forming the butterfly diagram in agreement
with the Parker-Yoshimura rule and solar-cycle observations. 
Unlike the flux-transport dynamo models, this model does not depend on the transport of magnetic field by meridional circulation at the bottom of the convection zone, and works well when the meridional circulation forms two cells in radius, as recently indicated by deep-focus time-distance helioseismology analysis of the SDO/HMI and SOHO/MDI data. We compare the new dynamo model with various characteristics if the solar magnetic cycles, including the cycle asymmetry (Waldmeier's relations) and magnetic `butterfly' diagrams.
\end{abstract}

\section{Basic Properties of Solar Magnetic Cycles}

\begin{figure}[t]
\begin{center}
\includegraphics[scale=0.65]{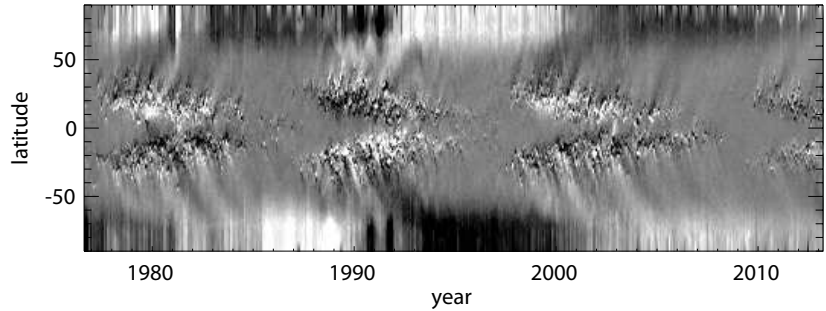}
\caption{Magnetic `butterfly' diagram shows the evolution of the
azymuthally averaged line-of-sight magnetic field a function of
latitude and time. The range of magnetic field strength is from
-10~G to +10~G. The synoptic magnetogram data are provided by the National Solar Observatory \label{fig1}}
\end{center}
\end{figure}

\begin{figure}[t]
\begin{center}
\includegraphics[scale=0.5]{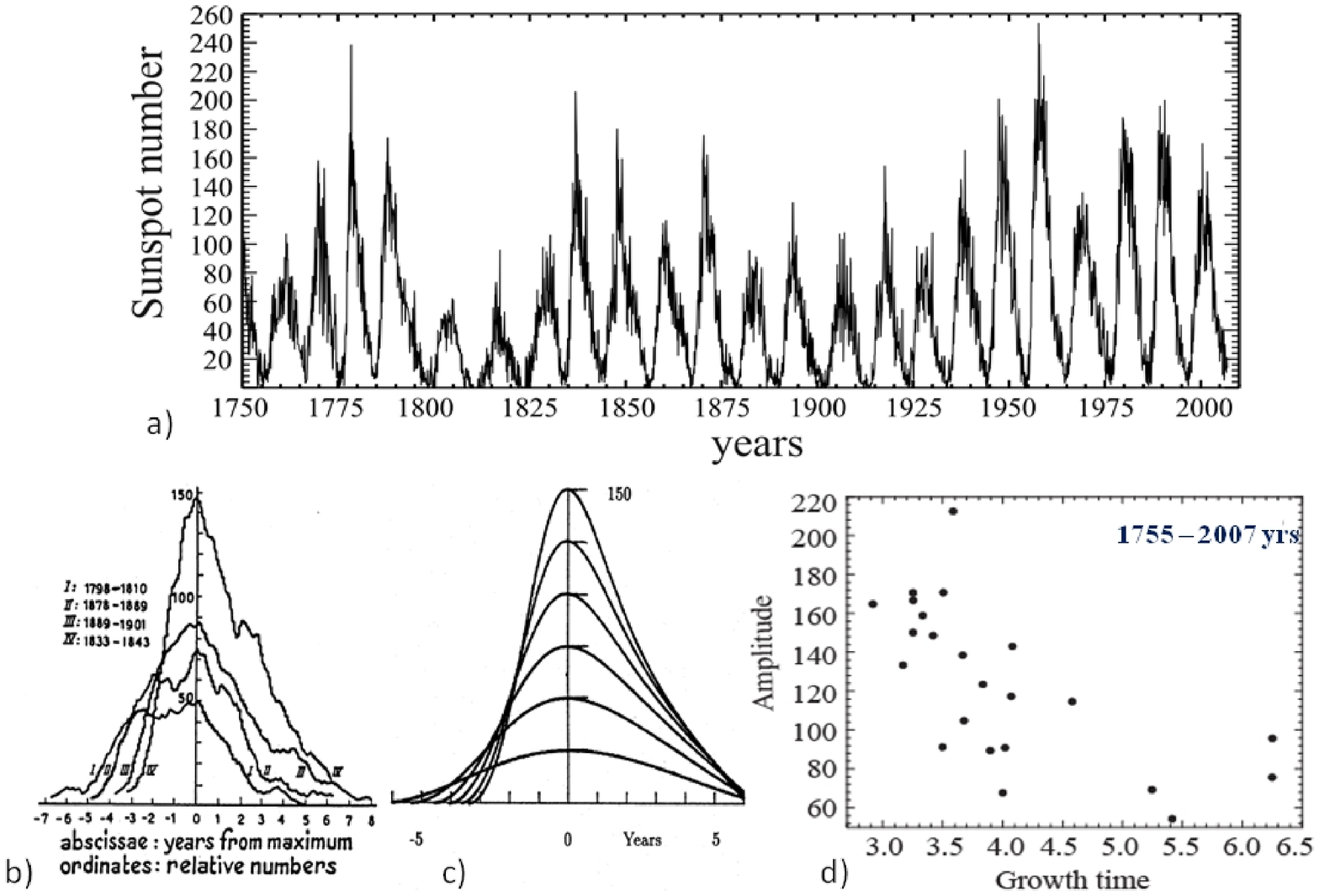}
\caption{a) The sunspot number as a function of time illustrates the
asymmetry of the solar cycles (the Waldmeier's effect); b)
comparison of the sunspot number curves for four cycles
\citep{Bracewell1988}; c) a model of the Waldmeier relations
\citep{Bracewell1988}; d) dependence of the cycle amplitude from the
growth time \citep{Kitiashvili2009}. \label{fig2}}
\end{center}
\end{figure}

The dynamo which operates in the solar convection zone and defines
the properties of sunspot cycles remains enigmatic despite
substantial observational and modeling efforts. In particular, the
solar dynamo models must explain the magnetic "butterfly" diagram
(Fig.~\ref{fig1}) and asymmetry of the sunspot cycles
(Fig.~\ref{fig2}a), so-called Waldmeier's effect
\citep{Waldmeier1935}. The magnetic butterfly diagram is obtained by
stacking azimuthally averaged synoptic magnetic field maps, provided
by the National Solar Observatory since 1976. 

The most prominent
features of the butterfly diagram are migration of the sunspot
formation zones towards the equator, migration of the following
polarity flux towards the poles, polarity reversals of polar magnetic
fields during sunspot maxima, and also polarity reversals of the toroidal magnetic field during the solar minima, which are observed as cyclic changes of polarity of leading and following sunspots in bipolar active regions (the Hale's law). The Waldweier's effect (non-linear
asymmetry of sunspot cycles) has the following three main
characteristics: 1) the sunspot number growth time is shorter than
the decay time (Fig.~\ref{fig2}b); 2) the growth time of strong
cycles is shorter than the growth time of weak cycles
(Fig.~\ref{fig2}c-d);  3) the strong cycles are shorter than the weak
cycles. The first property is often used for predicting the sunspot
maximum from the growth rate at the beginning of a cycle, using the
Waldmeier's "standard curves" (Fig.~\ref{fig2}c). Thus, the dynamo theories have to explain both the Hale's law and the Waldmeier's effect.

\section{Dynamo dilemma}

\cite{Bullard1955} suggested that the sunspot pairs represent parts of subsurface toroidal magnetic rings, emerged from the depth comparable with the size of these pairs, i.e. $~20$ Mm, and that the toroidal magnetic field is produced from the poloidal field by the latitudinal differential rotation beneath the solar surface. Generation of the toroidal magnetic field through stretching of the poloidal field by the differential rotation is a common feature of most solar dynamo models. However, the models differ in the depth of the toroidal field generation, and also in the physics of the reverse process of the poloidal magnetic field generation from the toroidal field.
\begin{figure}[t]
\begin{center}
\includegraphics[scale=0.5]{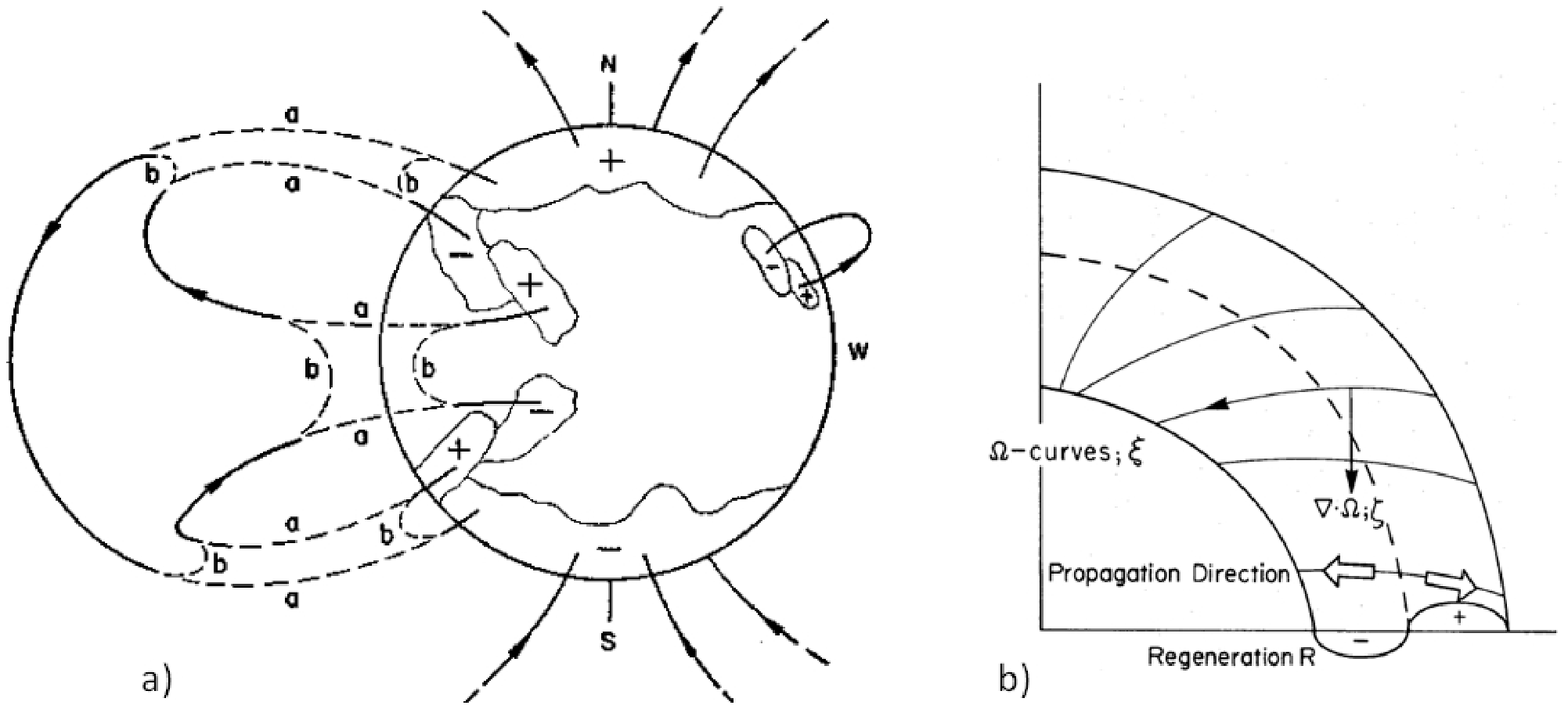}
\caption{ Dynamo dilemma: a) illustration of the Babcock-Leighton dynamo model \citep{Babcock1961,Leighton1969}: the toroidal magnetic field producing sunspot regions is generated by the differential rotation in the convection zone while the poloidal magnetic field is produced near the surface by magnetic flux diffusion, the equator-ward migration of sunspot formation zones is provided by the equator-ward meridional flow at the bottom of the convection zone \citep{Wang1989} ; b) illustration of the propagation direction of dynamo waves along the isorotation surfaces in the Parker-Yoshimura model \citep{Parker1955,Yoshimura1975}, the equator-ward migration of the sunspot zones requires  a decrease of the internal differential rotation rate towards the surface. \label{fig3}}
\end{center}
\end{figure}

\begin{figure}[t]
\begin{center}
\includegraphics[scale=0.5]{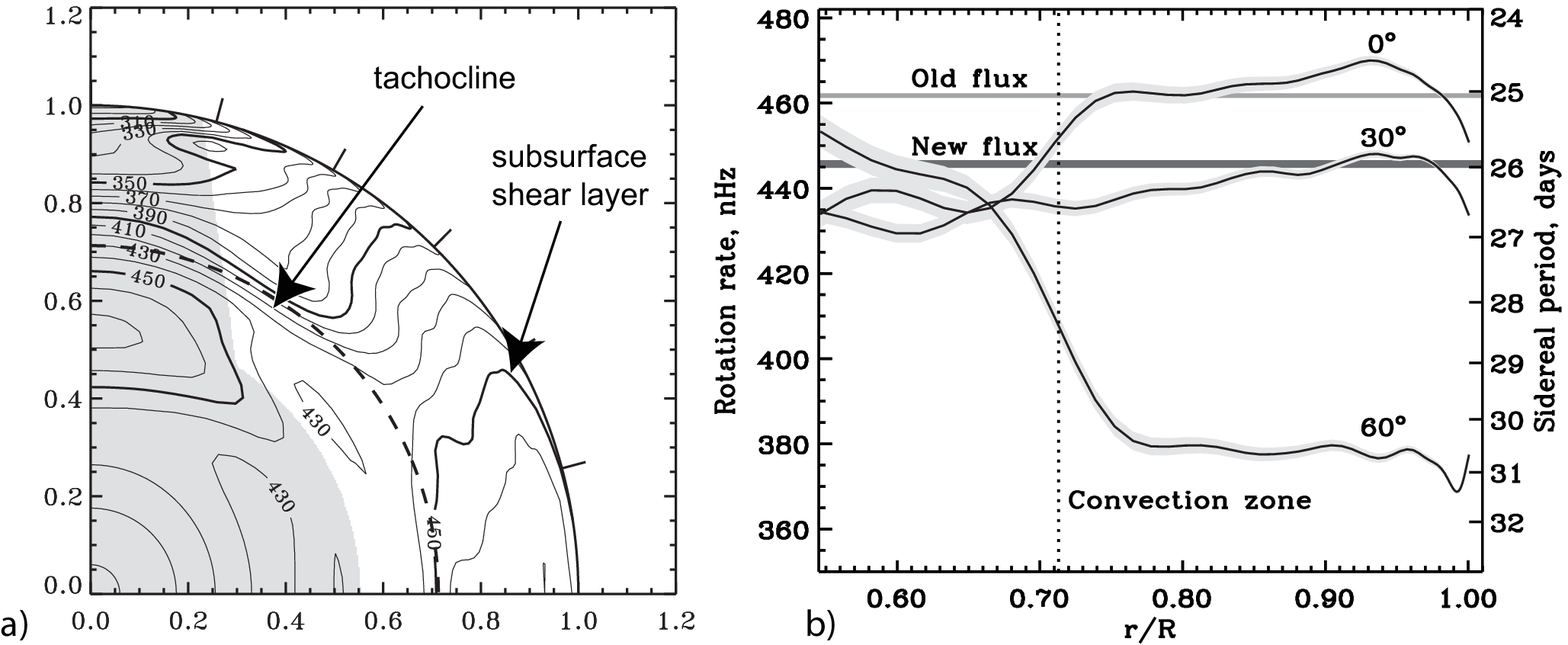}
\caption{a) Solar differential rotation inferred by helioseismology \citep{Schou1998}; b) comparison of the internal rotation rate at different latitudes with the rotation rate of emerging magnetic flux at the beginning of the solar cycles ("new flux") and at their end ("old flux") \citep{Benevolenskaya1999}.\label{fig4}}
\end{center}
\end{figure}

Two basic models of the solar dynamo operation have been developed. The first, so-called `flux-transport' model \citep{Babcock1961,Leighton1969} assumes that the poloidal magnetic field is produced by a combined action of the Coriolis force, turbulent diffusion and meridional circulation. In these models, the Coriolis force causes a tilt of emerging magnetic regions, relative to equator, which leads to preferential diffusion of magnetic flux of the trailing polarity and, subsequently, the polar field polarity reversals (Fig~\ref{fig3}a). The sunspot butterfly diagram is explained by the equator-ward meridional flow at the bottom of the convection zone, slowly transporting the regenerated toroidal magnetic field from mid latitudes towards the equator \citep{Wang1991}. Among the well-known difficulties of such models is the requirement to generate coherent magnetic flux tubes at the bottom of the convection zone with the field strength of $6\times 10^4-10^5$~G, the magnetic energy density of which substantially exceeds the turbulent energy density. In addition, helioseismology revealed that speed of the meridional circulation substantially varies with the solar cycle, mostly due to large-scale converging flows around active regions. These flows may significantly affect the polarward diffusion process.

In alternative models, initially suggested by \citet{Parker1955}, the poloidal field is generated by helical turbulence in the bulk of the convection zone, and is transported by turbulent diffusion in a form of `dynamo waves', travelling along the isorotation surfaces \citep[Fig.~\ref{fig3}b,][]{Yoshimura1975}. To explain the butterfly diagram this model requires that the rotation rate decreases towards the surface. However, measurements of the internal differential rotation by helioseismology showed that the rotation rate increases almost through the whole convection zone, except a shallow subsurface rotational shear layer, where the rotation sharply decreases (Fig.~\ref{fig4}). This means that the dynamo waves in the deep convection travel poleward, contrary to the sunspot butterfly diagram. This model, like the flux-transport model, could not explain the Waldmeier's effect.

Thus, both types of the solar dynamo models faced significant problems with explaining the solar magnetic cycles. \cite{Brandenburg2005} suggested that the dynamo process can be distributed in the convection zone but `shaped' in the subsurface layer, where the dynamo wave can migrate equatorward long the isorotation surfaces, according to the Parker-Yoshimura rule. This idea is supported the comparison of the magnetic flux rotation rate with the internal differential rotation \citep{Benevolenskaya1999}. This comparison showed that if the magnetic flux emerges radially and keep the rotation rate of its 'nest' then it most likely originates from the upper convection zone. If the magnetic flux tubes emerge from the base of the convection zone then the rotation rate of these flux tubes is generally slower than the observed rotation rate \citep{Weber2012}.

\section{Solar Dynamo Modeling - Paradigm Shift}

To investigate the idea of subsurface-shear-shaped dynamo suggested by
\citet{Brandenburg2005}, \cite{Pipin2011} calculated a mean-field model, in which the dynamo effect is
distributed in the bulk of the convection zone, and the toroidal
magnetic-field flux gets concentrated in regions of low turbulent diffusivity at the boundaries of the convection zone.
They showed that if the conditions at the top of the
convection zone such that the large-scale toroidal magnetic field
penetrates close to the surface, then the butterfly diagram for
the toroidal field in the upper convection zone is
formed by the subsurface rotational shear layer, following the Parker-Yoshimura rule, and that this can explain the observed equator-ward migration of the sunspot formation zone during the solar cycles. In previous dynamo models such penetration of toroidal field was prevented because of an artificially 
high turbulent diffusivity or/and because of the boundary conditions representing potential (vacuum) magnetic field outside the convection zone. It was shown that changing just one of these assumptions results in significant changes of the solar dynamo properties. In particular, the potential field assumption can be lifted by adding electrical conductivity in the top boundary condition \citep{Pipin2011}. Also, if the turbulent diffusivity is consistently calculated following the mixing-length theory the toroidal magnetic field penetrates into the subsurface shear layer even with the potential-field boundary condition \citep{Pipin2011b}. 

\begin{figure}[t]
\begin{center}
\includegraphics[scale=0.5]{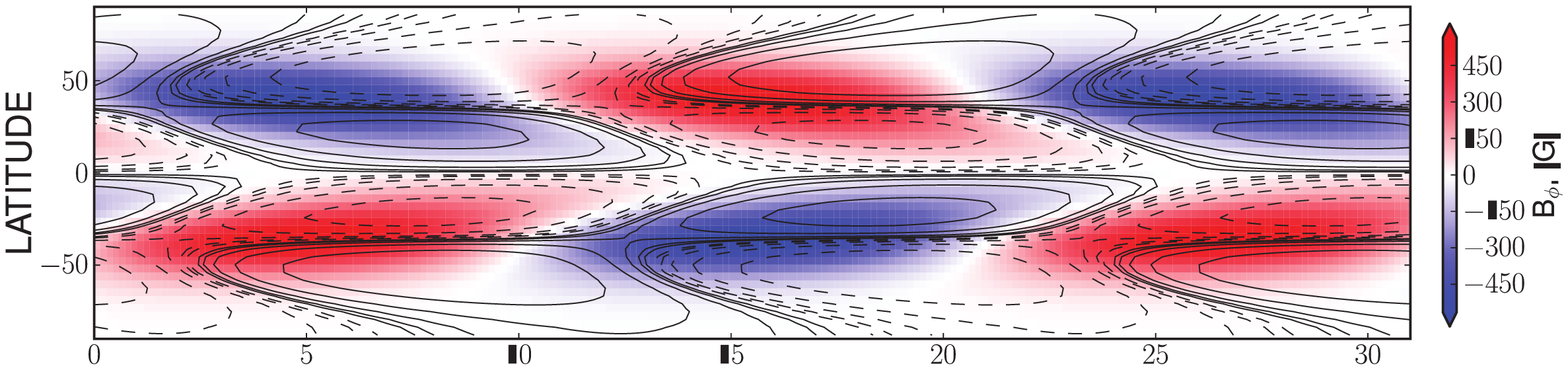}
\caption{Illustration of the dynamo model, which includes effects of the  near-surface rotational shear: the ``butterfly'' diagram of the toroidal magnetic field (color background) and the radial component of the poloidal field (contours) at the solar surface ($r=0.99~R_\odot$) \citep{Pipin2011}. \label{fig5a}}
\end{center}
%\end{figure}
%\begin{figure}[h]
\begin{center}
\includegraphics[scale=0.45]{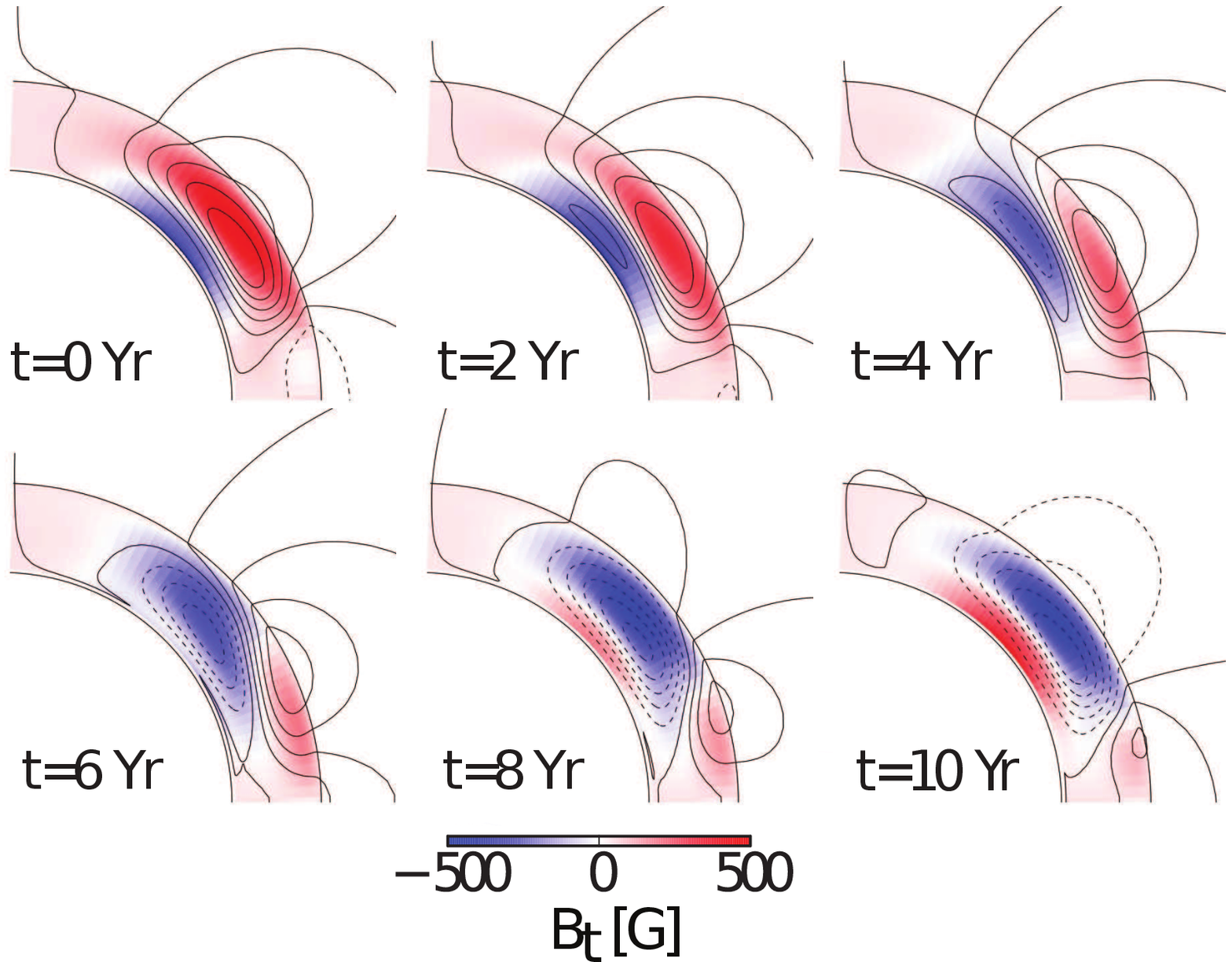}
\caption{Illustration of the dynamo model, which includes effects of the  near-surface rotational shear: the distribution of the toroidal (color) and poloidal magnetic field (contours) in the convection zone at different phases of the dynamo cycle \citep{Pipin2011}. \label{fig5b}}
\end{center}
\end{figure}

\begin{figure}[t]
\begin{center}
\includegraphics[scale=0.55]{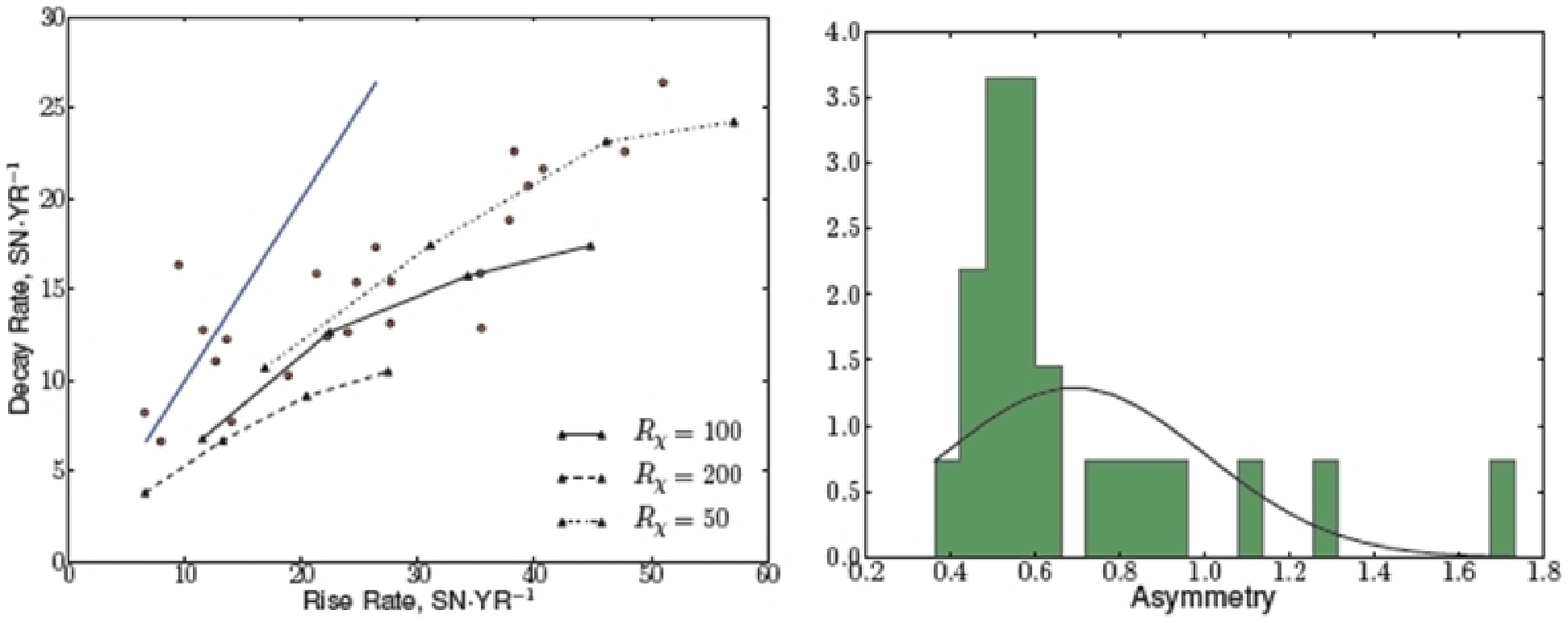}
\caption{Asymmetry of the sunspot cycles simulated in the dynamo model with the subsurface shear layer (Waldmeier's effect): a) the relationship between the rise and decay times for different values of a parameter $R_\xi$ that controls the dissipation rate of magnetic helicity; b) histogram of the asymmetry parameter  (the ratio of the rise and decay times) obtained from observations of the sunspot  cycles (Fig.~\ref{fig2}a) and from the model with $R_\xi=100$. \citep{Pipin2011b} \label{fig6}}
\end{center}
\end{figure}

The dynamo model with the subsurface rotational shear is illustrated in Figures~\ref{fig5a} and \ref{fig5b}, which show the butterfly diagram for the toroidal magnetic field and the corresponding evolution of the radial component of the poloidal field at the solar surface, and also the distribution of the toroidal and poloidal magnetic fields in the convection zone at different phases of the dynamo cycle. The butterfly diagram and the phase relation between the toroidal and poloidal magnetic field correspond quite well to the solar observations (Fig.~\ref{fig1}).

These results mean that the Parker's dynamo wave model can be consistent with both the helioseismology inferences and magnetic field data. This brings new life to this model and represent a paradigm shift in solar dynamo modeling. In this model with the subsurface shear the meridional circulation affects the dynamics of magnetic field, but it no longer plays a key role in the dynamo process. The period of magnetic cycles is determined by the speed of the dynamo waves, which is controlled by turbulent magnetic diffusivity, and the strength of the sunspot cycles is determined by the turbulent kinetic helicity. Both these quantities can be consistently estimated from turbulence models based on the mixing length theory. The observed variations among the solar cycles, particularly, in the cycle maxima can be explained by long-term fluctuations in the kinetic helicity \citep{Pipin2012}.

The new model can also explain the asymmetry of the sunspot cycles (the Waldmeier's effect), as a result of the magnetic helicity conservation which
provides dynamic quenching for the poloidal field generation (`alpha'-effect) \citep{Pipin2011b}. Figure~\ref{fig6} shows the relationship between the rise and decay times for three different values of a parameter $R_\xi$ which controls the dissipation rate of magnetic helicity (this parameter is similar to an effective turbulent magnetic Reynolds number), and a  histogram of the ratio of the rise and decay times obtained from the observations of the sunspot  cycles (Fig.~\ref{fig2}a) and from the model with $R_\xi=100$.

Our dynamo model has several advantages compared to the flux-transport models. In particular, it can explain the relative stability of the duration of sunspot cycles compared to the flux-transport models, for which the cycle duration depends on the speed of the meridional flows, which can vary substantially during the cycles \citep{Zhao2004}. It does not require the formation of compact toroidal magnetic flux tubes with the field strength of $\sim 60-100$ kG, which are problematic because the energy density of which exceeds the turbulent energy equipartition level, and also does require to explain coherent emergence of these flux tubes from the bottom of the convection zone. In our model, the magnetic field of sunspots and active regions can be provided by the magnetic field emerging from relatively shallow layers, similar to the process simulated numerically by \citet{Stein2012}. However, the tilt of bipolar active regions (the Joy's law) is not yet explained.
%It may be related to anisotropy of magnetic diffusivity caused by the differential rotation.

\section{New Helioseismology Observations and Constraints
on Dynamo Models}

\begin{figure}[t]
\begin{center}
\includegraphics[scale=0.4]{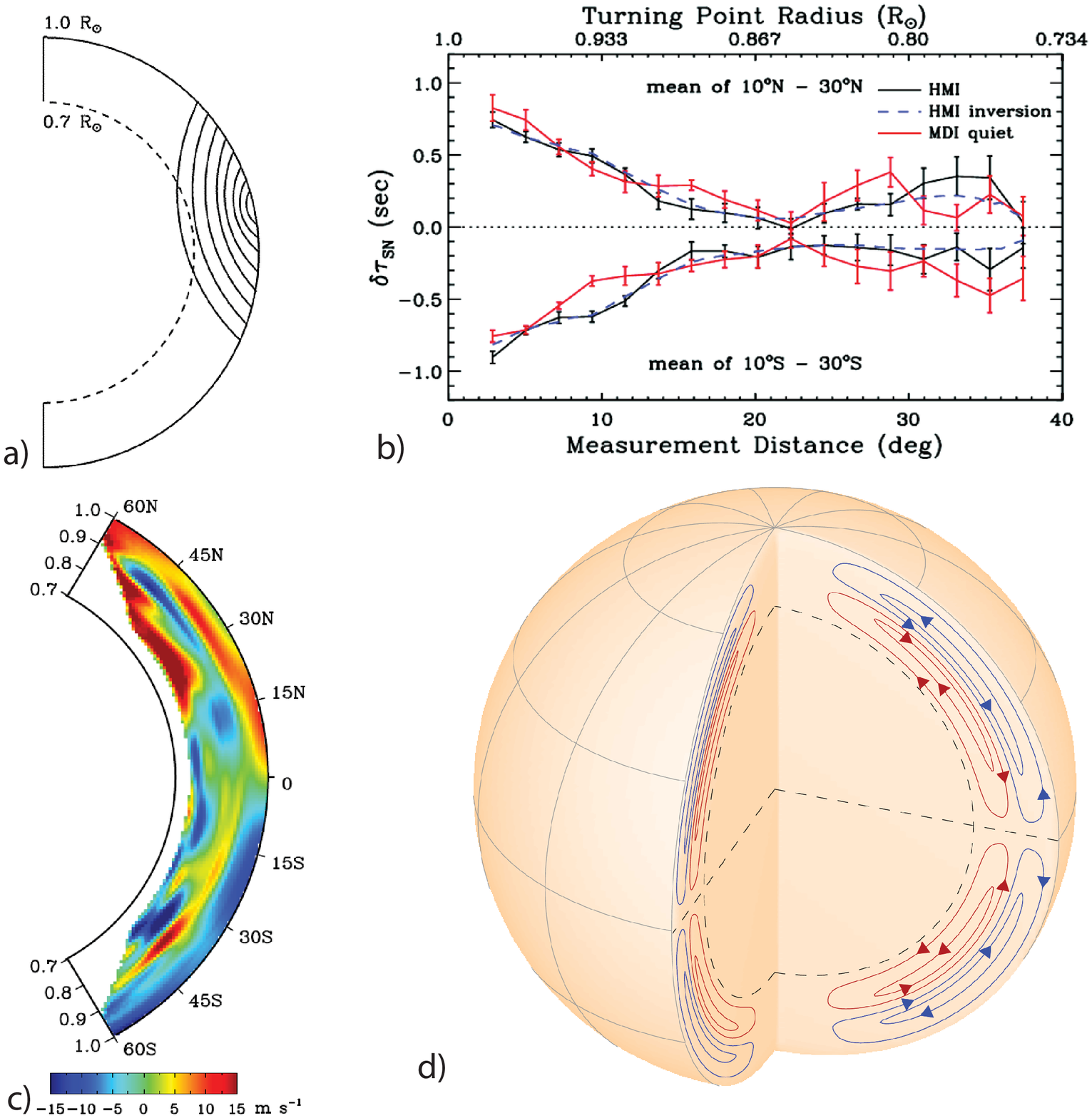}
\caption{Helioseismic measurements of the meridional flow: a) the deep-focus time-distance helioseismology measurement scheme; b) variations of acoustic travel times, due to the meriodional flows, as a function of the wave travel distance or the radius of the wave lower turning point (positive values are for the measurements in the northern hemisphere, and negative values are for the southern hemisphere), the solid curves are the measurement results (red curve is for the SOHO/MDI instrument, black curve is for the SDO/HMI), the dashed curve shows the travel times calculated from the inversion results; c) the distribution of the meridional flow speed obtained by inversion of the SDO/HMI travel time measurements; d) schematic illustration of the double-cell meridional circulation \citep{Zhao2013}.  \label{fig7}}
\end{center}
\end{figure}

In addition to the differential rotation, the knowledge of the internal meridional flows is critical for developing the solar dynamo models. It was long assumed that the meridional flows are represented in the Northern and Southern hemispheres by circulation cells occupying the whole convection zone, and that at the flow is directed from the equator to the poles at the top of the convection zone and is reversed at the bottom. Since the flow speed is quite low ($\sim 20$ m/s) it is very difficult to measure these flows by helioseismology. Previous time-distance measurements by \citet{Giles1997} established that the polar-ward meridional flow extends into deeper convection zone, but were not able to detect the return meridional flow. The initial evidence that the return meridional flow can be shallow, starting at depth 20 Mm was obtained by \citet{Braun1998}. Later \citet{Mitra-Kraev2007} obtained an estimate of the depth of the flow reversal at around 40 Mm. In both cases, the error estimate was comparable with the flow signal itself, and also no verification and testing of their techniques was done. 
The high-resolution helioseismology data from the Solar Dynamics Observatory (SDO) HMI instrument provides new opportunities for improving measurements of the meridional flow. Important advantages of the SDO/HMI data over the previous helioseismology observations are that the data have much higher resolution and almost uninterrupted, and also that the solar oscillations are observed not only in the Doppler shift, but also in other spectral parameters: continuum intensity, line width and line depth \citep{Scherrer2012}. 

Using these multi-parameter measurements \citet{Zhao2012b} found previously unknown center-to-limb systematic shifts of the acoustic travel times. These systematic shifts are different for the different observables, and probably caused by a combination of the wave leakage and line formation effects in the solar atmosphere, due to changes of the effective observing height from the center to the limb. Near the limb the spectral line is formed higher in the atmosphere than near the center. This may cause systematic travel time shifts of the order of few seconds \citep{Nagashima2012}, which have to be taken into account in measurements of the meridional flows. The exact mechanism of these shifts is not understood, and \citet{Zhao2012b} developed an empirical correction procedure. 

In this procedure
the systematic variations were determined by measuring the travel-time variations along the equatorial regions during the periods when the solar rotation axis is perpendicular to the line of sight. After the substraction of the center-to-limb variations the measurements from the different observables gave very similar results. Inversion of the corrected travel times showed that the previous helioseismic measurements overestimated the meridional flow speed by about 10 m/sec. The corrected speed is more consistent with the surface flow speed obtained directly from the Doppler shift \citep[e.g.][]{Hathaway2012}.

Recently, \citet{Zhao2013} used a deep-focus measurement scheme (Fig.~\ref{fig7}a) to measure the meridional flow speed in the deep convection zone. They used the HMI Doppler-shift data covering its first 2-year period from 2010 May 1 through 2012 April 30, and calculated the time-distance cross-covariance functions for 60 measurement distances ranging from $\sim 2$ to 44 degrees, covering almost the whole depth of the convection zone, according to the acoustic ray theory. The cross-correlation functions were averaged for the same latitudes, and then averaged again over one-month intervals. The acoustic travel times were determined by fitting the Gabor wavelet functions \citep{Kosovichev1997}, and corrected for the center-to-limb variation. Finally, for each distance the travel times were averaged again over the whole 2-year period, and used to determine the meridional flow speed by inversion in the acoustic ray-path approximation \citep{Kosovichev1996}. 

Figure~\ref{fig7}b shows the the North-South travel time differences as a function of the wave travel distance or the radius of the wave lower turning point (positive values are for the measurements in the northern hemisphere, and negative values are for the southern hemisphere), the solid curves are the measurement results (red curve is for the SOHO/MDI instrument, black curve is for the SDO/HMI). In the asymptotic ray-path approximation, the sensitivity of acoustic travel times to the internal flows depends on the local sound speed and the angle between the ray path and the flow velocity. Therefore, most of the travel time sensitivity comes from the near-surface layers, where the sound-speed is low and from the region around the lower turning point of the acoustic ray paths because the waves travel along the flow. The sensitivity to the near-surface region is dominant, so that the travel times do not change sign when the waves travel through the deep regions of return flows. The return flow effects are reflected in the rate of decrease of the travel times with the depth of the wave turning point, or equivalently with the increase of the travel distance. 

The travel times from the SDO/HMI and also from the SOHO/MDI shown in Fig.~\ref{fig7}b indicate a rapid decrease for the waves traveling into the deep convection zone, which is steeper than predicted for the standard single-cell models of the meridional circulation with the return flow near the bottom of the convection zone. Such rapid decrease indicates that the return flow is rather shallow. Also, quite unexpectedly, the travel times start rising for the acoustic waves traveling through the lower half of the convection zone, indicating the existence of deep poleward flows or a secondary circulation cell. The inversion results shown in Fig.~\ref{fig7}c confirm this \citep{Zhao2013}. Thus, the new helioseismology measurements based on very long time series of observations of solar oscillations provide a strong evidence that the meridional circulation in the solar convection zone has a complicated structure and consists of at least two circulation cells stacked along the radius, as schematically illustrated in Figure~\ref{fig7}d. Such double-cell meridional flow may be consistent the pole-ward migration of supergranulation, as recently suggested by \citet{Hathaway2012}.

This result immediately puts in question the standard flux-transport dynamo models, which have to rely on the single-cell meridional flow to carry the magnetic flux towards the equator at the bottom of the convection zone in order to explain the sunspot butterfly diagram. This result also raises the question about how this type of meridional circulation can affect the distributed dynamo model with the subsurface shear layer.

\section{Dynamo model with double-cell meridional circulation}

\begin{figure}[t]
\begin{center}
\includegraphics[scale=0.5]{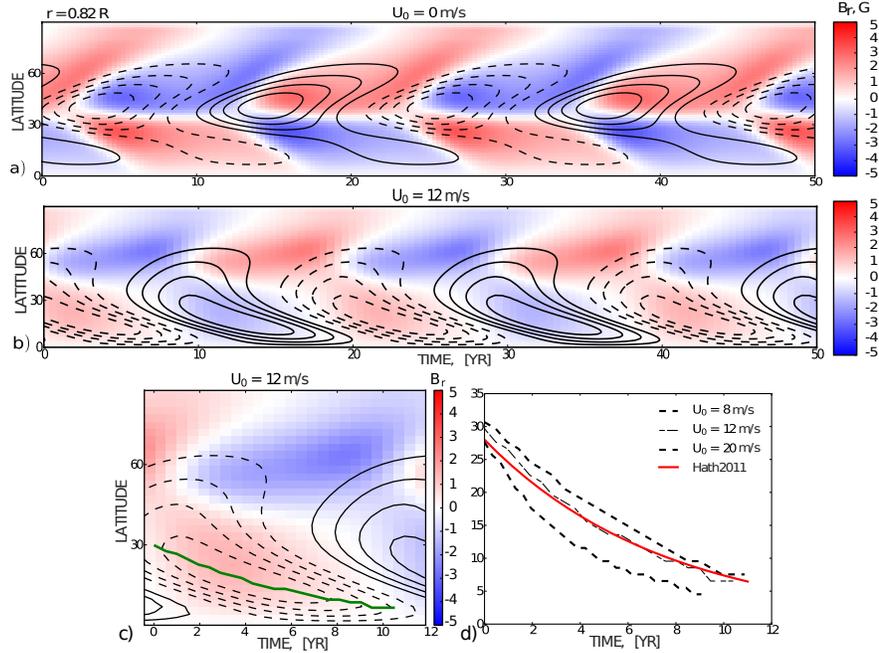}
\caption{Evolution of the large-scale magnetic field inside the convection zone for the
dynamo model with the meridional circulation speed $U_0 = 12$ m/s. The field lines show of the poloidal component of the mean magnetic field, and the toroidal magnetic field which varies in the range $\pm 0.6$ kG) is shown by the background images \citep{Pipin2013}.\label{fig8}}
\end{center}
\end{figure}

\begin{figure}[t]
\begin{center}
\includegraphics[scale=0.46]{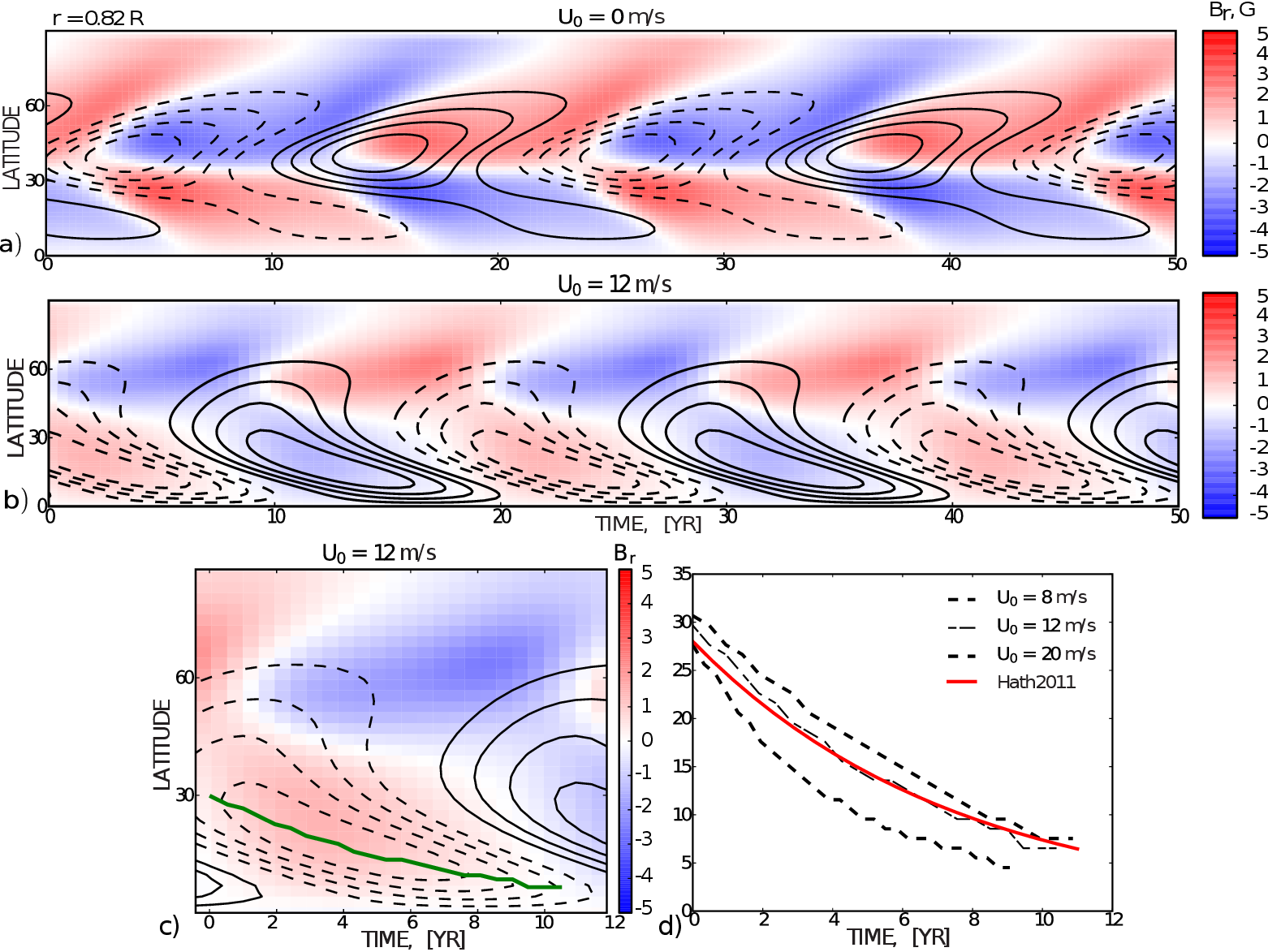}
\caption{ The magnetic ``butterfly'' diagrams at $r = 0.82~R$ for: a) the dynamo model without the meridional circulation; b) for the model with the double-cell meridional circulation of $U_0 = 12$ m/s; c) tracking of the toroidal field maximum, which is an assumed latitudinal zone of sunspot formation; d) comparison of the toroidal field maxima for the models with different characteristic meridional flow speed with the observed speed of sunspot zone migration \citep{Hathaway2011}. The toroidal field
is shown by contours (plotted for $\pm 100$~G range), and the surface radial magnetic field is shown by background images. We draw these diagrams only for one hemisphere because  the antisymmetric mode (dipole-like) is dominant \citep{Pipin2013}.\label{fig9}}
\end{center}
\end{figure}

To investigate the effects of the double-cell meridional circulation we calculated dynamo models with such circulation, using the mean-field magnetohydrodynamics theory approach which includes detailed modeling of the mean electromotive force and turbulent diffusion coefficient in the so-called
``minimal tau-approximation'' \citep[e.g.][]{Pipin2008}. 
The tau-approximation suggests that  the second-order correlations 
do not vary significantly on the timescale $\tau_c$ that corresponds 
to the typical turnover time of
the convective flows. The theoretical calculations are done for the
anelastic turbulent flows, and  take into account the effects of
density stratification, spatial inhomogeneity of the intensity  of
turbulent flows and inhomogeneity of the large-scale magnetic
fields. The effects of the large-scale inhomogeneity of the turbulent
flows and magnetic fields are calculated to the first order of the
Taylor expansion in terms of the ratio typical
spatial scales of turbulence and the
mean quantities \citep[for further details, see][]{Pipin2008}. 
The meridional circulation is modeled in the form two spherical-shape circulation cells along the radius, occupying the whole convection zone \citep{Pipin2013}, as shown in Figure~\ref{fig7}d.

Results for this dynamo model are shown in Figures~\ref{fig8} and \ref{fig9}. Contrary to previous flux-transport dynamo models, which fail for such type of the meridional circulation \citep{Jouve2007}, it is found that the dynamo model can robustly reproduce the basic properties of the solar magnetic cycles for a wide range of model parameters and the circulation speed. The best agreement with observations
is achieved when the surface speed of meridional circulation is about 12
m/s. For this circulation speed the simulated sunspot activity shows the good
synchronization with the polar magnetic fields. 

The toroidal magnetic field of the new cycle is generated near the bottom of the convection zone by the differential rotation. Simultaneously, in Fig.~\ref{fig8}, we see a start of generation of the poloidal magnetic field (contour lines). The dynamo wave propagates by a turbulent diffusion process almost radially to the surface following the Parker-Yoshimura rule \citep{Parker1955,Yoshimura1975}. However, the propagation of the wave is inclined to the equator because
of the anisotropy of the turbulent diffusion and turbulent transport effects. Near the surface the turbulent downward pumping and the subsurface rotational
shear stop the radial propagation and deflect the dynamo wave
toward the equator. The near-surface meridional circulation and the turbulent diffusion bring the decaying poloidal field to the poles.
The meridional circulation modifies the propagation of the dynamo wave. 
It is found that the toroidal magnetic field is involved in clockwise advection by the bottom circulation cell in a manner similar to the flux transport models. Near the surface the poloidal field migrates towards the poles at high latitudes and towards the equator at low latitudes.

Figure~\ref{fig9} shows the time-latitude ``butterfly'' diagrams of the toroidal (contours) and radial (background image) magnetic fields evolution in the upper part of the solar convection zone 
for the meridional flow speed $U_0 = 0$ and 12 m/s. In both cases (with and without the circulation) there is a qualitative agreement with observations. However, the maximum of the toroidal magnetic field migrates closer to the equator for the model with the circulation. Also, in this case the butterfly diagram wings are wider in
latitude than in the case without circulation. Also it is found that the double-cell circulation reduces
the latitudinal width of the polar branch for the radial magnetic field evolution and also reduces the overlap between the cycles.

\section{Conclusion}

The helioseismology discoveries of the rotational subsurface shear layer and the double-cell meridional circulation require to re-examine dynamo models of the
solar activity cycles. They lead a new paradigm of the solar dynamo distributed in the convection zone and equator-ward migrating dynamo waves in the sub-surface shear layer. The double-cell meridional circulation affects the dynamics of the large-scale magnetic fields, and puts additional constraints. In particular, if the observed equator-ward migration of the sunspot formation zone (`butterfly' diagram) is linked to migration of toroidal magnetic field in the convection zone then the magnetic field of sunspots may emerge from the depth of about 120~Mm. This is generally consistent with the observed rotation rate of surface magnetic fields of active regions and the internal differential rotation determined by helioseismology. Further helioseismology investigation will shed more light on the solar dynamo mechanism. 

\acknowledgements The authors would like to thank the Fujihara Foundation of Science for support. This work was supported by the NASA LWS grants NNX09AJ85G to UCLA, and NNX09AT36G to Stanford University.

%\bibliography{dynamo_v1}

\end{document}